\documentclass[12pt,pra,reprint,superscriptaddress,amsmath,amsfonts,amssymb]{revtex4-1}

\usepackage[pretty]{revquantum}
\usepackage[english]{babel}
\usepackage{graphicx}
\usepackage[caption=false]{subfig}
\usepackage{mathrsfs}
\usepackage{amsmath}
\usepackage{amsthm}
\usepackage{thmtools, thm-restate}
\usepackage{bm}
\usepackage{xcolor}
\usepackage{hyperref}
\usepackage{hypernat}
\usepackage[utf8]{inputenc}
\usepackage[T1]{fontenc}
\usepackage{cleveref}
\usepackage{soul}
\usepackage{dsfont}
\usepackage{algorithm}
\usepackage{algpseudocode}

\begin{document}
\title{Analysis of Neural Network Predictions for Entanglement Self-Catalysis}

\author{Tha\' is M. Ac\' acio}
\affiliation{YELP, 9 Appold Street, 5th floor, London, EC2A 2AP, United Kingdom}
\author{Cristhiano Duarte}
\email[Corresponding author: ]{c.silva1@leeds.ac.uk}
\affiliation{School of Physics and Astronomy, University of Leeds, Leeds LS2 9JT, United Kingdom}
\affiliation{International Institute of Physics, Federal University of Rio Grande do Norte, 59070-405 Natal, Brazil}
\affiliation{Wigner Research Centre for Physics, H-1121, Budapest Hungary}

\date{\today}
\begin{abstract}

Machine learning techniques have been successfully applied to classifying an extensive range of phenomena in quantum theory. From detecting quantum phase transitions to identifying Bell non-locality, it has been established that classical machines can learn genuine quantum features via classical data. Quantum entanglement is one of the uniquely quantum phenomena in that range, as it has been shown that neural networks can be used to classify different types of entanglement. Our work builds on this topic. We investigate whether distinct models of neural networks can learn how to detect catalysis and self-catalysis of entanglement. Additionally, we also study whether a trained machine can detect another related phenomenon - which we dub transfer knowledge. As we build our models from scratch, besides making all the codes available, we can study a whole gamut of paradigmatic measures, including accuracy, execution time, training time, bias in the training data set and so on.

\end{abstract}

\maketitle

\section{Introduction}\label{Sec.Introduction}

Quantum physics exhibits striking phenomena that do not find a counterpart in the classical world. From its foundational level to its more informational aspect, classically counter-intuitive examples abound in the quantum world. The possibility of having non-local correlations~\cite{Bell66}, contextuality~\cite{KS67}, genuine superposition~\cite{Zurek03}, the impossibility of cloning or broadcasting information~\cite{HBMA07}, the fact that quantum events may be  observer-dependent~\cite{Cavalcanti21} and non-absolute~\cite{MNDC21}, are just a few of the puzzling features displayed by quantum theory. Among those, we can safely say quantum entanglement is one of the most prominent, well-studied and popular non-classical phenomenon that to date may still shock newcomers and more experienced researchers alike.

Mathematically, a multipartite system is said to be entangled when it cannot be decomposed as a convex combination of its individual constituents~\cite{NC00,HHHH09}. However, more than a mathematical object, entanglement has assumed over the past decades a key role in quantum computation~\cite{BennettEtlAl97}, quantum communication~\cite{BennettEtAl93,ZhangEtAl06}, quantum cryptography~\cite{MinderEtAl19,BB14} and more recently in quantum networks~\cite{BritoEtAl21}. It is hands-down an important resource~\cite{WS98}. Nonetheless, it is a fragile resource, as quantum systems tend to decohere and lose their quantum properties over time when interacting with the surrounding environment~\cite{Zurek09,BKZ06}. Therefore, as an important but hard to handle resource, it is of utter importance to study how to manipulate it. Protocols involving transformations of entanglement abound, but deterministic local operations and classical communications (LOCC) are the golden standard when exploring entanglement transformations mathematically~\cite{ChitambarElAl14} - mainly due to the inherent symmetry imposed by the scenario.

Entanglement manipulation is a remarkable task, as it reveals another subtlety of the quantum world. There are pairs of quantum states that are not interconvertible but that become one-way convertible in the presence of an extra source that is not consumed over the course of the process. That situation has been dubbed the catalysis of entanglement~\cite{JP99}. As a matter of fact, it has been recently discovered that entanglement transformations are even more curious, as there are pairs of interconvertible quantum states that become one-way convertible in the presence of redundant information - a situation that received the name of self-catalysis, insofar the redundant information is not consumed during the conversion~\cite{DDT16}.

Although the study of entanglement conversion is critical, as it points out how to obtain certain resources from others, mathematically speaking, it is a rather cumbersome and costly task~\cite{BFBS18,DF01,JP99}. In this sense, any new methodology that can speed up the process of classification is extremely welcome. Even if such methodology comes with the downside of
introducing small errors, that would not be a big compromise, as entanglement transformation is robust to small perturbations~\cite{DDT16}. It is in this sense that we see classical algorithms of machine learning as the perfect toolbox to address this task. In this work, in particular, we will be making extensive use of neural networks, as they are a toolbox built to identify patterns from a dataset and can, therefore, be trained to classify whether a particular fed-in transformation happens to hold true.

The main aim of this contribution is twofold. First, we want to provide a proper answer for whether classical algorithms of machine learning are able to classify pair of vectors that are convertible, catalytic-convertible, and self-catalytic-convertible. This would help to answer if classical machines, as a proxy for classical agents, can learn about the quantum world through classical data. Second, we want to give an in-depth analysis of every portion of our model, at least as in-depth and didactically as possible. We will open the usually kept closed black-boxes, and show why we have chosen such-and-such models, why we have selected certain parameters, why the lack of accuracy is not a problem and the like. We will also touch on some caveats of our own models. To facilitate the reading, we also make available all the codes and data we are using in this work ( Appendices~\ref{Sec.AppGraphs} and~\ref{Sec.AppCodes}).

This work is divided as follows. Sections~\ref{Sec.CatalysisAndSelfCatalysis} and~\ref{Sec.Methodology} are introductory. In the former, we present an abridged overview of entanglement transformation, focusing primarily on self-catalysis. In the latter, we discuss our methodology, from how we generate our training dataset to why we have chosen a particular network model. Section~\ref{Sec.Results} summarises our main results. Finally, we conclude the paper with sec.~\ref{Sec.Conclusion}, where we review our findings, establish connections with other works in the literature and point out further investigations. The subsequent appendices contain supplementary material, mainly on neural networks, useful to newcomers.

\section{Entanglement Transformation \\ Catalysis and Self-Catalysis}\label{Sec.CatalysisAndSelfCatalysis}

\subsection{Deterministic Transformations}\label{SubSec.Transformations}

As we mentioned in the introduction, entanglement is more than a mere mathematical property of physical systems described by quantum theory. Over the past decades, quantum entanglement has acquired the status of crucial ingredient in many informational, computational, and even foundational tasks. Entanglement can be used to improve communication protocols, establish physically secure communication networks and used as a tool to investigate the split between the classical and quantum realm. Put another way, we have learned that entanglement should be seen as a resource, and treating it like so, it is of utmost importance to know how to harness it, transform it, and maintain it intact - see ref.~\cite{HHHH09} for an extensive review and ref.~\cite{CG19} for a more contemporary approach. An efficient framework to analyse deterministic entanglement transformations is the goal of this work. We will focus on transformations involving pure bipartite states.

Mathematically, given two finite-dimensional Hilbert spaces $\mathcal{H}_{A}$ and $\mathcal{H}_{B}$, we say that a quantum system represented by $\rho^{AB} \in \mathcal{D}(\mathcal{H}_{A}) \otimes \mathcal{D}(\mathcal{H}_{B})$ is \emph{separable} when it can be writen down as:
\begin{equation}
    \rho^{AB} = \sum_{a=1}^{K_{A}}\sum_{b=1}^{K_{B}}p_{ab}\rho_{a}^{A} \otimes \rho_{b}^{B},
    \label{Eq.DefSeparableState}
\end{equation}
where $0 \leq p_{ab} \leq 1$ with $\sum_{ab}p_{ab}=1$ , and $\{\rho_{a}^{A}\}_{a}$ and $\{\rho_{b}^{B}\}_{b}$ are families of positive semidefinite trace-class operators acting on $\mathcal{H}_{A}$ and $\mathcal{H}_{B}$ respectively. When such a decomposition is not possible, we say that $\rho^{AB}$ is \emph{entangled}~\cite{NC00}. 

In plain words, eq.~\eqref{Eq.DefSeparableState} says that a certain state $\rho^{AB}$ is separable whenever it can be cast as a (convex) combination of product states. In this sense, separable states have a clear division between their constituent parts - even though they might be classically correlated via $\{p_{ab}\}_{ab}$. Entangled states, on the other hand, are those states in which a classically crystal clear concept of parts starts to fade away - as if the whole were more than the sum of its parts~\cite{Duarte16}. In fact, it is exactly that characteristic that grants entanglement the status of a useful resource. 

From a resource theoretical point of view, deciding whether or not a given state $\rho^{AB}$ is entangled is a crucial problem. Nonetheless, the task of finding necessary and sufficient criteria to determine the nature of the correlations shown by multipartite states is not an easy exercise~\cite{HHH96}. For bipartite states of low dimensions, it is known how to determine whether or not a given state is entangled with a simple criterion~\cite{Peres96}. For higher dimensions, the problem gets richer and more complicated. The characterisation of multipartite entanglement is an area of intense and deep research to date~\cite{NevenEtAl21,SHKS20}. 

Still considering entanglement as a resource, it is natural to explore how to transform one type of entanglement into another. The paradigmatic scenario is that of a seller that only trades a determined family of entangled states, such that it is on the buyer to transform the bought state into something useful for them. Imagine the hypothetical situation where the seller has access and sells only maximally entangled states. In this case, a buyer who knows how to extract entanglement may ask whether it is worth purchasing states from this seller - and in the affirmative case, how many? If we focus on an emblematic class of transformations involving bipartite states, questions of this type have a precise mathematical description, and this is exactly what we approach in the next paragraphs.

The emblematic class of transformations we mentioned above are named \emph{local operations with classical communications (LOCC)}. They are predominant in the study of bipartite entanglement as they encompass what we judge natural when dealing with states shared between two distant parties. Broadly speaking, the LOCC paradigm is a two-party communication protocol, involving many rounds of information exchange between the parties, such that (a) on every round, each agent is allowed to operate locally on their laboratory; (b) the classical output coming out of these operations can be classically communicated between the agents; (c) on the next round of local operations, those operations may be conditioned on the bit of information each party has received. We refer to ref.~\cite{ChitambarElAl14} for an in-depth and detailed study of LOCC transformations. For the present work, (a)-(c) suffices, for it is the neat mathematical characterisation arising from LOCC~\cite{Nielsen99} that we will centre our attention at. 

\begin{theorem}[Nielsen's Theorem]
Let $\ket{\psi}$ and $\ket{\phi}$ be two bipartite states in $\mathcal{H}_{A} \otimes \mathcal{H}_{B}$. Transforming $\ket{\psi}$ into $\ket{\phi}$ is possible via local operations with classical communication if and only if the following inequalities hold true:
\begin{align}
    \sum_{i=1}^{l}\lambda^{\psi}_{i} \leq \sum_{i=1}^{l}\lambda^{\phi}_{i}, 
\label{Eq.ThmNielsenMajorizationIneqs}
\end{align}
for every l, where $\lambda_{\psi}$ is the Schmidt vector of $\ket{\psi}$ whose entries are listed in a non-increasing ordering. More succinctly, 
\begin{align}
 \ket{\psi} \rightarrow \ket{\phi}  \mbox{ iff } \lambda_{\psi}  \preceq \lambda_{\phi},    
\end{align}
where $\preceq$ means the majorization order defined by inequalities in~\eqref{Eq.ThmNielsenMajorizationIneqs}.
\label{Thm:NielsenThm}
\end{theorem}

In conclusion, the manipulation of bipartite entanglement is mathematically well-characterised. The possibility of conversion $\ket{\psi}$ into $\ket{\phi}$ can be promptly determined from checking a list of inequalities involving their Schmidt vectors $\lambda_{\psi}$ and $\lambda_{\phi}$. Entanglement transformation reduces to a particular comparison between probability vectors - losing no generality, we are free to focus our attention exclusively on probability vectors.

\subsection{Examples of Entanglement Transformations}\label{SubSec.ExamplesCatalysisAndSelf}

The last section concludes by saying that the study of transformations involving bipartite pure states is equivalent to studying majorization. Put another way, to study entanglement transformation is to study the majorization preorder defined over $\mathbb{V}^{d}$ the set of probability vectors with $d$ entries:
\begin{align}
    \mathbb{V}^{d}:=\left\lbrace(p_1,...,p_d) \in \mathbb{R}^{d}; \,\,  0 \leq p_{i} \leq 1 \mbox{ and } \sum_{i=1}^{d}p_i=1  \right\rbrace
    \label{Eq.DefVd}
\end{align}

In this more simplified framework, seen as a preorder on $\mathbb{V}^{d}$, entanglement transformation via LOCC possesses two crucial and surprising features. The examples below illustrate these two core ideas.

\textit{\textbf{Example 1:} Suppose that dim$(\mathcal{H}_{A})=d=$ dim$(\mathcal{H}_{B})$. Maximally entangled states $\ket{\psi} \in \mathcal{H}_{A} \otimes \mathcal{H}_{B}$ have, in this case, the same Schmidt decomposition given by
\begin{align}
\lambda_{\psi}=\left( \frac{1}{d},\frac{1}{d},...,\frac{1}{d} \right).
\label{Eq.ExMaximallyEntangled}
\end{align}
Remarkably, $\lambda_{\psi}$ is a minimal element for the majorization preorder. In other words, any other probability vector $(p_1,p_2,...,p_d) \in \mathbb{V}^{d}$ with entries listed in a non-increasing order must satisfy:
\begin{align}
    \lambda_{\psi} \preceq (p_1,p_2,...,p_d).
    \label{Eq.ExMaxEntangledIsMinimal}
\end{align}
Eq.~\eqref{Eq.ExMaxEntangledIsMinimal} is a direct consequence of the fact that $\lambda_{\psi}=(1/d,1/d,...,1/d)$ is the barycenter of $\mathbb{V}^{d}$~\cite{MOA11}. Similarly, any product state $\ket{\phi} in \in \mathcal{H}_{A} \otimes \mathcal{H}_{B}$ has Schmidt vector given by $\lambda_{\phi}=(1,0,0,...,0)$ and satisfies:
\begin{align}
    (p_1,p_2,...,p_d) \preceq \lambda_{\phi}.
    \label{Eq.ExProductStatesMaximal}
\end{align}
Which means that product states act as maximal elements for the majorization preorder.} 

\textit{\textbf{Example 2:} Let $\ket{\alpha}$ and $\ket{\beta}$ be two vectors in $\mathcal{H}_{A} \otimes \mathcal{H}_{B}$ whose Schmidt decomposition are respectively
\begin{equation}
 \lambda_{\alpha}=\left(\frac{1}{2},\frac{1}{4},\frac{1}{4},0\right) \mbox{ and } \lambda_{\beta}=\left(\frac{2}{5},\frac{2}{5},\frac{1}{10},\frac{1}{10}\right).
 \label{Eq.ExSchmidtVectorsIncomparable}
\end{equation}
In this case, it holds that $\lambda_{\alpha} \npreceq \lambda_{\beta}$ and $\lambda_{\beta} \npreceq \lambda_{\alpha}$. It suffices to note that 0.5 > 0.4 and that 0.75 < 0.8, so that the first two inequalities in eq.~\eqref{Eq.ThmNielsenMajorizationIneqs} do not hold true regardless of the comparison one is trying to establish.} 

In plain words, the first example shows that maximally entangled states act as a sort of universal resource as it may be transformed into any other pure state. Additionally, the example also shows that pure product states can be obtained out of any other entangled state via LOCC. The second example says that majorization is not a trivial order over $\mathbb{V}^{d}$, as there are probability vectors that that are not comparable - and consequently, entangled states that are not interconvertible. The focus of this work is precisely on these situations of non-interconvertibility, for such non-interconvertibility, as we argue below, may not be the end of the story.

\subsection{Catalysis and Self-Catalysis}\label{SubSec.CatalysisAndSelf}

In the last section, we mention that non-interconvertibility (example 2) may not be the end of the story for entanglement transformations. As put forward in ref.~\cite{JP99}, it can be the case that not-allowed transformations end up happening in the presence of an additional source of entanglement - that is not consumed during the process. Entanglement transformations turned possible by an extra resource that is not destroyed throughout the process are called \emph{catalytic transformations}. The example below revisits the example 2 and makes this discussion more precise. 

\textit{\textbf{Example 3:} Let $\ket{\alpha}$ and $\ket{\beta}$ be two vectors in $\mathcal{H}_{A} \otimes \mathcal{H}_{B}$ whose Schmidt decomposition are respectively
\begin{equation}
 \lambda_{\alpha}=\left(\frac{1}{2},\frac{1}{4},\frac{1}{4},0\right) \mbox{ and } \lambda_{\beta}=\left(\frac{2}{5},\frac{2}{5},\frac{1}{10},\frac{1}{10}\right).
 \label{Eq.ExSchmidtVectorsIncomparable}
\end{equation}
as we have seen, neither $\lambda_{\alpha} \preceq \lambda_{\beta}$ nor $\lambda_{\beta} \preceq \lambda_{\alpha}$. Now, consider an additional entangled state $\ket{\gamma}$ whose Schmidt decomposition is given by $\lambda_{\gamma}=(6/10,4/10)$. Tensor product of two states is translated as the Kronecker product of the respective Schmidt vectors, so that:
\begin{align}
     &\lambda_{\alpha \otimes \gamma}=\left(\frac{24}{100}, \frac{24}{100}, \frac{16}{100}, \frac{16}{100}, \frac{6}{100}, \frac{6}{100}, \frac{4}{100}, \frac{4}{100} \right) \nonumber \\ 
     \mbox{ and } \nonumber \\
     &\lambda_{\beta \otimes \gamma}=\left( \frac{30}{100}, \frac{20}{100}, \frac{15}{100}, \frac{15}{100}, \frac{10}{100}, \frac{10}{100}, 0, 0 \right).
    \label{Eq.CatalysisForAlphaBetaGamma}
\end{align}
One may run the comparisons of eq.~\eqref{Eq.ThmNielsenMajorizationIneqs} and check that $\lambda_{\alpha \otimes \gamma} \preceq \lambda_{\beta \otimes \gamma}$. In this case, although the ordinary transformation  $\ket{\alpha} \rightarrow \ket{\beta}$ does not hold true, the presence of an additional and not-consumed resource $\ket{\gamma}$ allows for $\ket{\alpha} \otimes \ket{\gamma} \rightarrow \ket{\beta} \otimes \ket{\gamma}$ - note that the extra state returns intact after the transformation takes place.    
}

The fact that an additional state may boost entanglement transformations is remarkably by itself. Curiously enough, maximally entangled states cannot act as catalysts for any transformation whatsoever - which may be counterintuitive at first sight, as we are used to considering maximally entangled states as the ultimate resource. This follows from the fact that maximally entangled states have a uniform Schmidt decomposition, so that if two states are not-interconvertible, multiplying their Schmidt coefficients by the same number does not affect their original majorization relation.

Even more astonishingly, redundant information may also act as a catalyst for forbidden entanglement transformations. This scenario was thoroughly investigated in ref.~\cite{DDT16} where deterministic entanglement transformations made possible via redundant information were dubbed \emph{self-catalytic} transformations. The following example explores this point. 

\textit{\textbf{Example 4:} Let $\ket{\alpha},\ket{\beta}$ and $\ket{\alpha'},\ket{\beta'}$ be two pairs vectors in $\mathcal{H}_{A} \otimes \mathcal{H}_{B}$ whose Schmidt decomposition are respectively $\lambda_{\alpha}=\left(0.900, 0.081,0.010,0.009\right)$ and  $\lambda_{\beta}=\left(0.950, 0.030, 0.020, 0\right)$ for the first pair, and 
 $\lambda_{\alpha'}=\left(0.928, 0.060, 0.006, 0.006\right)$ and $\lambda_{\beta'}=\left(0.950, 0.030, 0.020\right)$ for the second pair. In this case, neither of the ordinary transformations hold true: $\ket{\alpha} \nleftrightarrow \ket{\beta}$ as well as $\ket{\alpha'} \nleftrightarrow \ket{\beta'}$. For the first pair of states, one piece of redundant information is already sufficient for the transformation to hold
 \begin{align}
 \ket{\alpha} \otimes \ket{\alpha} \rightarrow \ket{\beta} \otimes \ket{\alpha},
 \end{align} 
as $\lambda_{\alpha \otimes \alpha} \preceq \lambda_{\beta \otimes \alpha}$. Something more remarkably happens to the second pair, as only after we have plugged in six copies of $\ket{\alpha}$ into the mix is that the transformation turns out to be true:
 \begin{align}
 \ket{\alpha} \otimes \ket{\alpha}^{\otimes 6} \rightarrow \ket{\beta} \otimes \ket{\alpha}^{\otimes 6}.
 \end{align} 
 }

Before concluding, we should comment that resource theories with catalysts are not an exclusivity of quantum entanglement~\cite{GourEtAl15,BrandaoEtAl15}. In fact, catalytic assisted transformations are not a privilege of quantum theory~\cite{LY01} either, and have been thoroughly studied~\cite{Fritz17,CFS16} ever since the publication of D. Jonathan and M. Plenio's seminal paper~\cite{JP99}.

All in all, deterministic transformations of entanglement via local operations with classical communication is a rich and complex phenomenon. Even those transformations that at first sight may seem forbidden can become possible when additional resources are used but not consumed - which may be crucial for practical purposes. Although Thm.~\ref{Thm:NielsenThm} gives a useful mathematical criterion to check whether a particular transformation holds, it is simply impracticable in practice  - numerically and analytically. Our work comes precisely to fill this gap. We investigate whether a machine can learn how to classify entanglement and if differently trained machines can detect different transformations. We will focus on different network configurations, different optimisers, different strategies of training, and on the effect of measure concentration, or bias, of the dataset. The figure of merit is the accuracy of each machine learning algorithm (and also the execution time).  In the next section, we explain in more detail our methodology.

\section{Methodology}\label{Sec.Methodology}

\subsection{Our Chosen Model}\label{SubSec.OurNNModel}

\begin{figure}
    \centering
    \includegraphics[scale=0.3]{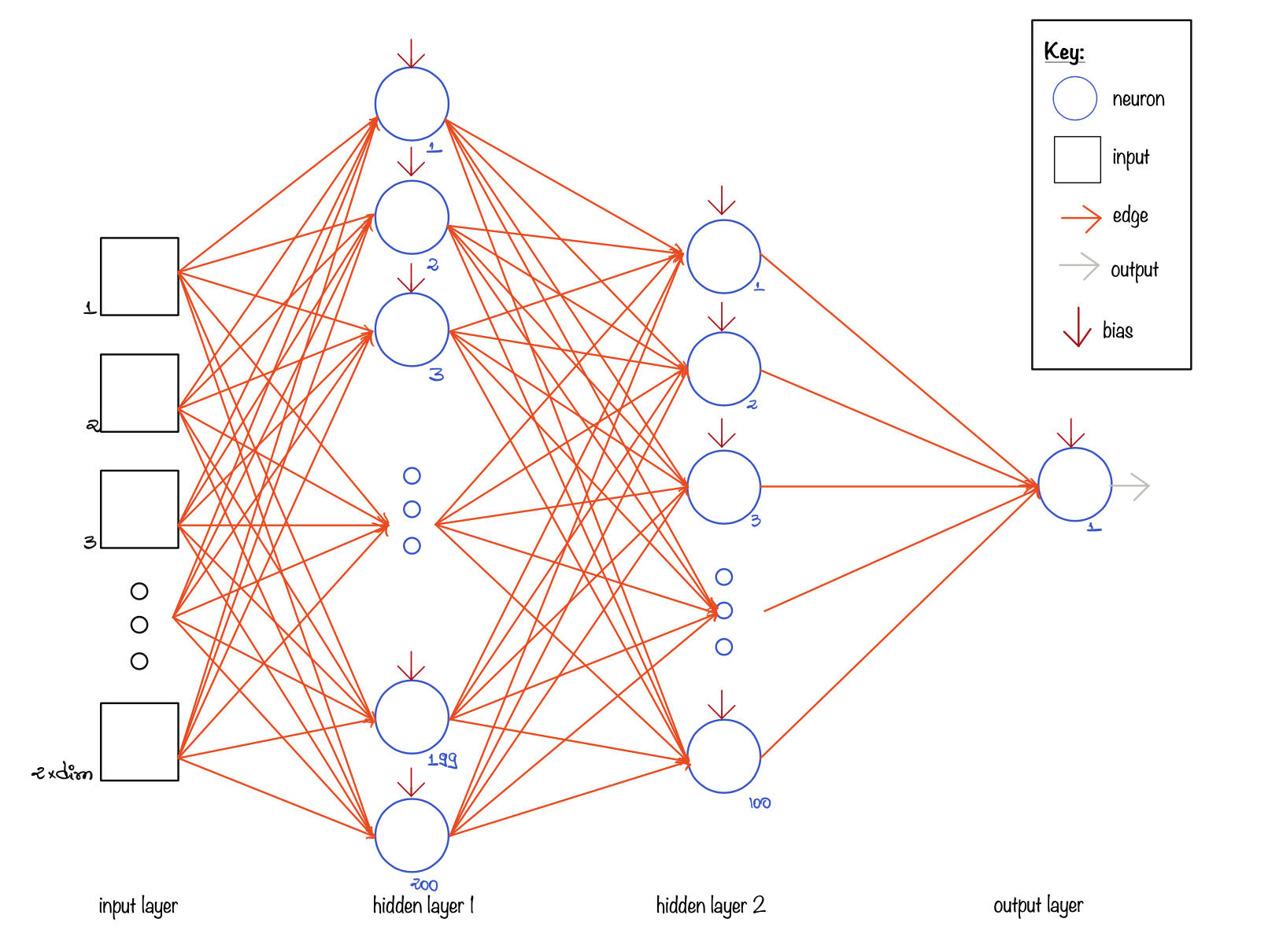}
    \caption{Schematic representation of our chosen model (colours on-line).}
    \label{Fig.NetworkModelConfiguration}
\end{figure}

In this section we discuss the neural network model we are working with. Among many others features, we have focused on test accuracy, test loss, training duration, and the total duration of the test phase in order to decide upon a particular model. These measures, or features, are crucial as they shown the potential computational gains one would get when switching from standard numeric investigations~\cite{DDT16} to machine learning techniques to studying self-catalysis. We will conclude that we can already get a high accuracy even if we work with a small network containing just a few layers.  
\begin{table}
  \includegraphics[scale=0.19]{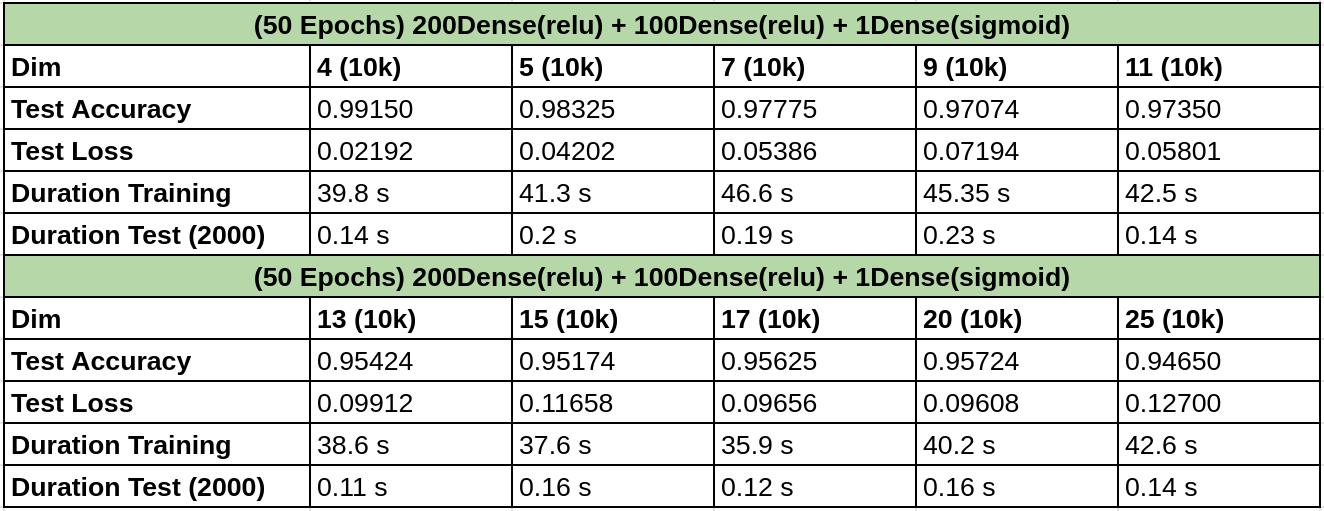}
  \caption{Chosen neural network model. The table displays how the test accuracy, test loss, training duration and test duration vary across different dimensions. The training dataset contains 8000 samples, while the test dataset contains 2000}
   \label{Fig_Table_Our_Model}
\end{table}

Our model consists of a network with two hidden layers and one output layer - see fig.~\ref{Fig.NetworkModelConfiguration}. The first hidden layer has 200 neurons, the middle layer has 100 neurons and the final output layer has a single neuron. In each hidden layer we have used a rectified linear unit activation function also known as \emph{relu}, and for the output layer we have chosen a sigmoid distribution as the activation function - see appendix~\ref{Sec.AppIntroduction} for a gentle introduction to these terms. As we are classifying the behaviour of pairs of probability vectors in a certain simplex $\mathbb{V}^{d}$, for each  dimension $d$ we have generated one million samples (\emph{i.e.} one million pairs of probability vectors) and divided them into two groups. 4/5 of these vectors were used to train the model, and the remaining 1/5 were left to the test phase. Finally, we have thoroughly explored a substantial number of training epochs, but we have stopped at 50 epochs, as this was the number of training epochs that brought the higher accuracy without compromising the duration of the training stage - see the data we provide in appendix~\ref{Sec.AppGraphs}.

To enrich our analysis, we have also investigated the role played by different optimisers. Although the definition of a well-suited optimiser is usually part of the model choice, we have decided to leave this door open and analyse the effects that different optimisers have on our classification problem - this is to show the extent to which variations in the algorithm can lead to significantly different results.

Table~\ref{Fig_Table_Our_Model} summarises the main aspects of the neural network model we are working with. For the sake of comparison, in the files made available in Appendix~\ref{Sec.AppGraphs} we are displaying other concurrent models that we have also considered in our study. In comparison with other models, our chosen model performs better in almost all considered measures - in particular it has also outperformed other networks when considering larger datasets, see table~\ref{Fig_Table_Our_Model_OneMillion} and the tables also displayed in the material made available in appendix~\ref{Sec.AppGraphs}.

\begin{table}
  \includegraphics[scale=0.23]{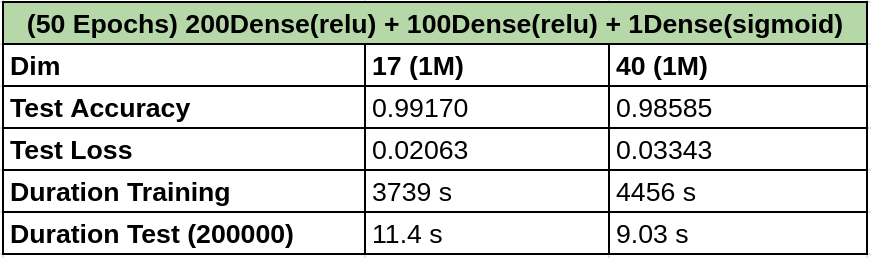}
  \caption{Chosen neural network model. The table displays how the test accuracy, test loss, training duration and test duration vary across different dimensions for a total of one million data points}
   \label{Fig_Table_Our_Model_OneMillion}
\end{table}

\subsection{Data Generation}\label{SubSec.DataGeneration}

The training of machine learning algorithms strongly relies on datasets provided by an external source to the models. In particular, supervised classification models identify patterns on that dataset and learn how to categorise each data-point into one of the known classes. The quality of the dataset directly impacts the outputs of the model. For this reason, it is crucially important to have a representative and fair dataset that contains all the classes to be learned by the model - preferentially in a balanced way. This section summarises how we have obtained our training data.

The pseudocode displayed in algorithm \ref{alg:generate_samples} explains the process of how we generate our data-points~\footnote{The complete code used for generate the samples can be found on GitHub \cite{AD21}}. Given $dim$ the desired dimension, and $n$ the number of pairs of vectors, the algorithm generates two sets $\{\alpha_1,...,\alpha_n\}$ and $\{\beta_1,...,\beta_n\}$ of $n$ normalised random vectors of size $dim$, following the uniform probability distribution, and whose entries are sorted in a non-increasing order. 

Then, using eqs.~\eqref{Eq.ThmNielsenMajorizationIneqs}, we check whether the majorization relation holds true across all the pairs of vectors in this set. In other words, we check whether $\alpha_i \preceq \beta_j$ as well as $\beta_j \preceq \alpha_i$ hold for all $i$ and $j$. The Boolean output resulting from the comparison together with the each compared pair are written in an external file that will be used to train the ML algorithm in the next step.

\begin{figure}
    \centering
    \includegraphics[scale=0.6]{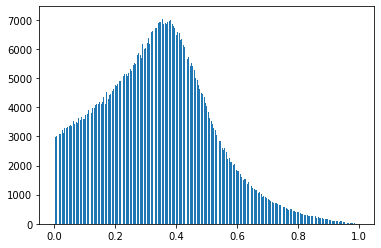}
    \caption{Histogram - Concentration of measure for $10^6$ vectors of $dim=3$. Our data generation method privileges small entries}
    \label{Fig.HistogramConcentrationOfMeasureD3}
\end{figure}

\begin{figure}
    \centering
    \includegraphics[scale=0.6]{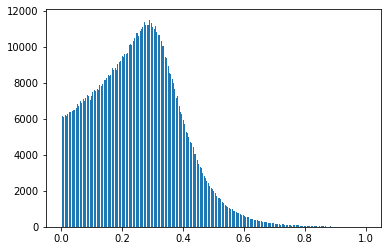}
    \caption{Histogram - Concentration of measure for $10^6$ vectors of $dim=4$. Our data generation method privileges small entries}
    \label{Fig.HistogramConcentrationOfMeasureD4}
\end{figure}

As we mention in the above paragraph, it is crucial to have a balanced and fair dataset, covering all the classes to be classified. Although we have found pairs of vectors falling in both classes we want to classify (True or False for majorization), our generation method is skewed and shows a high concentration around small entries - the same behaviour was also studied and detected in ref.~\cite{DDT16}. Fig.~\ref{Fig.HistogramConcentrationOfMeasureD3} and fig.~\ref{Fig.HistogramConcentrationOfMeasureD4} display that behaviour for smaller dimensions. For larger dimensions we refer to \cite{AD21}. In this sense, our findings are only valid within this bias. Nonetheless, we firmly believe that our conclusions would not dramatically change if we had chosen a different sample method.

\section{Results}\label{Sec.Results}

We dedicate this section to our main results. We first address whether classical machines can learn how to detect entanglement transformations. Put another way, whether one can train machines to check majorization relation with high accuracy. Second, we present our results on whether machines specialized in detecting entanglement can also be used to detect self-catalysis. Finally, we investigate a hybrid approach using the usual majorization formula and an appropriate machine to decide on higher-order self-catalysis. 

\subsection{Machines Learn Majorization - Entanglement Classification}\label{SubSec.ResultsMajorization}

We wanted to determine whether classical machine learning algorithms learn how to classify entanglement transformations. Via Nielsen's theorem (thm.~\ref{Thm:NielsenThm}), we have translated this problem into a majorization comparison between two probability vectors in $\mathbb{V}^{d}$. In this sense, one may say that we are equivalently studying whether classical machine learning algorithms learn the majorization pre-order. 

\begin{figure}
  \includegraphics[scale=0.35]{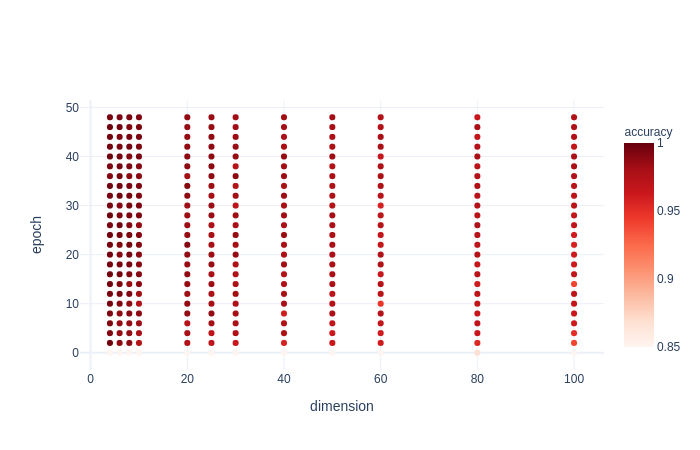}
  \caption{Majorization accuracy through epochs for different dimensions - optimiser: adam.}
   \label{Fig_Majorization_Adam}
\end{figure}

\begin{figure}
  \includegraphics[scale=0.35]{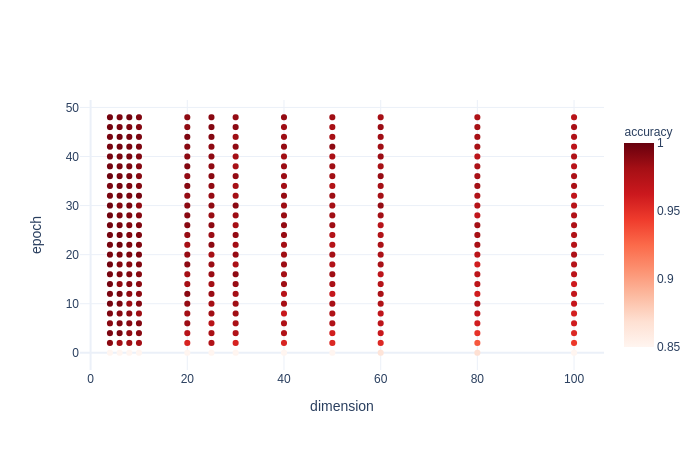}
  \caption{Majorization accuracy through epochs for different dimensions - optimiser: adadelta.}
   \label{Fig_Majorization_Adadelta}
\end{figure}

\begin{figure}
  \includegraphics[scale=0.35]{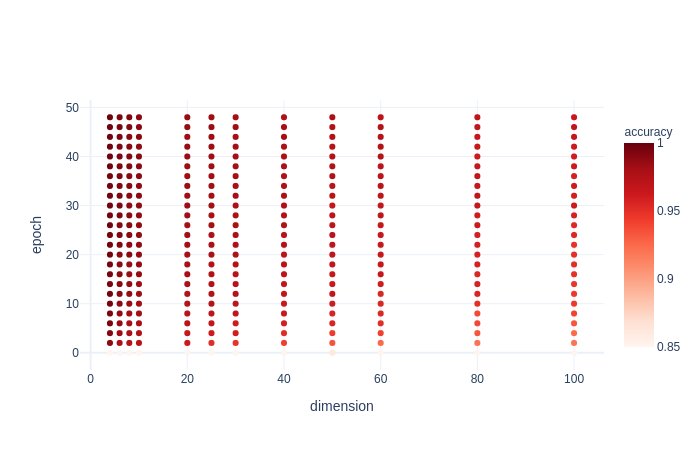}
  \caption{Majorization accuracy through epochs for different dimensions - optimiser: adagrad.}
   \label{Fig_Majorization_Adagrad}
\end{figure}

\begin{figure}
  \includegraphics[scale=0.35]{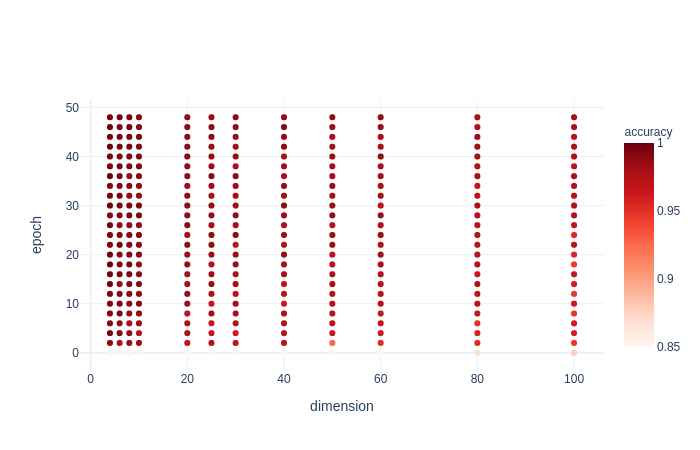}
  \caption{Majorization accuracy through epochs for different dimensions - optimiser: rmsprop.}
   \label{Fig_Majorization_Rmsprop}
\end{figure}

\begin{figure}
  \includegraphics[scale=0.35]{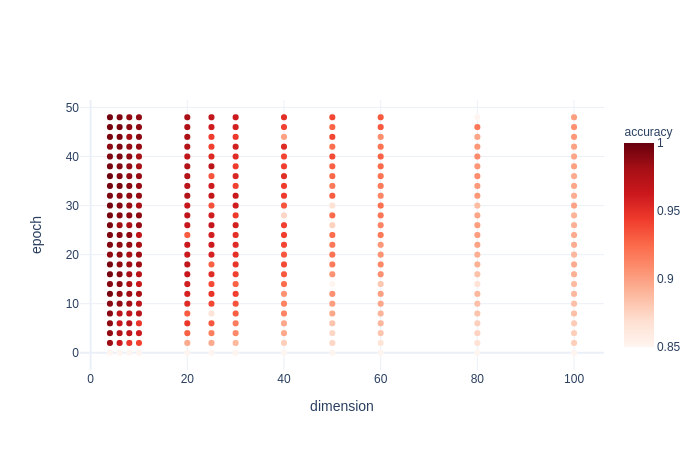}
  \caption{Majorization accuracy through epochs for different dimensions - optimiser: sgd.}
   \label{Fig_Majorization_Sgd}
\end{figure}

\begin{figure}
  \includegraphics[scale=0.35]{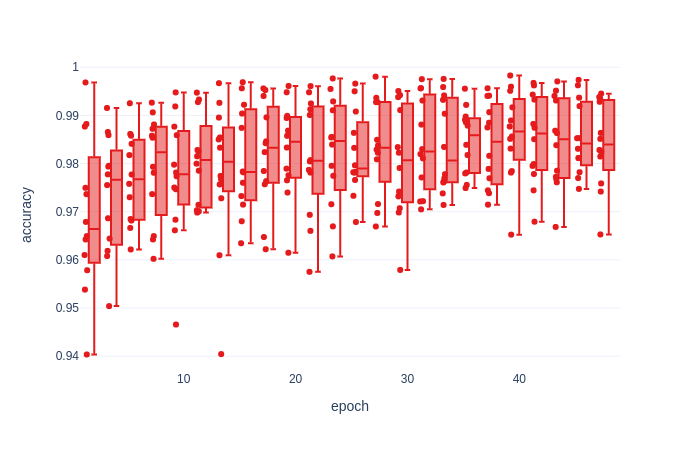}
  \caption{Box-plot showing the variance of the accuracy by epoch - optimiser: adam. Each point represents a dimension.}
   \label{Fig_Boxplot_Adam}
\end{figure}

\begin{figure}
  \includegraphics[scale=0.35]{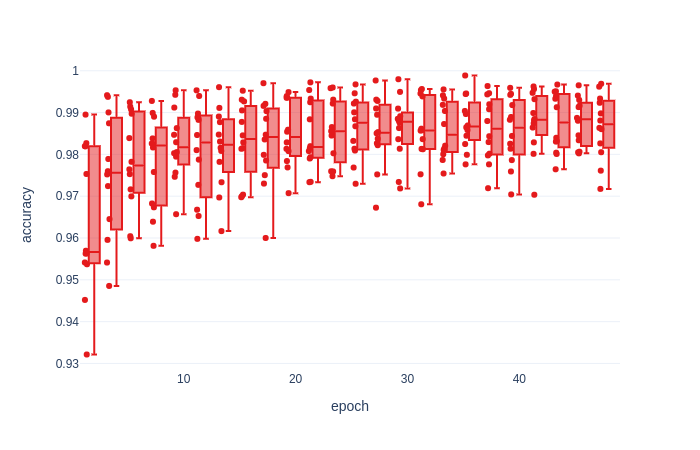}
  \caption{Box-plot showing the variance of the accuracy by epoch - optimiser: adadelta. Each point represents a dimension.}
   \label{Fig_Boxplot_Adadelta}
\end{figure}

\begin{figure}
  \includegraphics[scale=0.35]{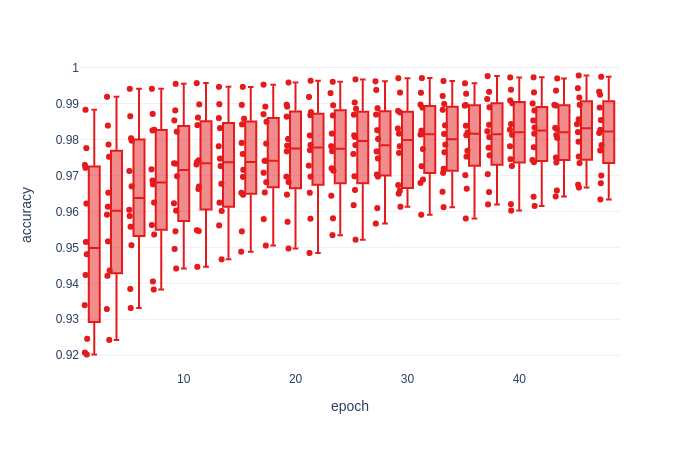}
  \caption{Box-plot showing the variance of the accuracy by epoch - optimiser: adagrad. Each point represents a dimension.}
   \label{Fig_Boxplot_Adagrad}
\end{figure}

\begin{figure}
  \includegraphics[scale=0.35]{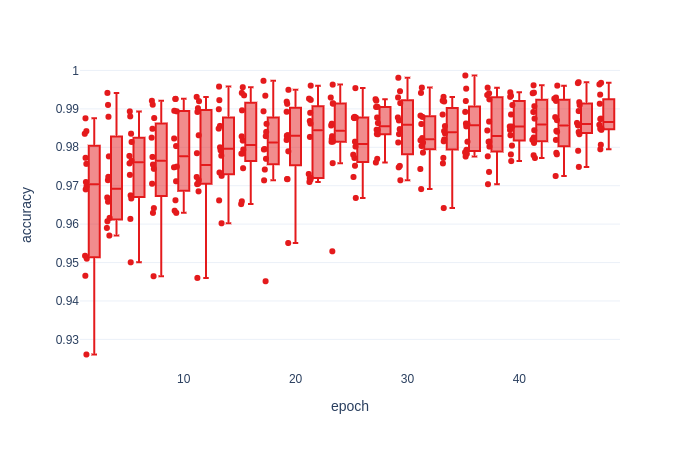}
  \caption{Box-plot showing the variance of the accuracy by epoch - optimiser: rmsprop. Each point represents a dimension.}
   \label{Fig_Boxplot_Rmsprop}
\end{figure}

\begin{figure}
  \includegraphics[scale=0.35]{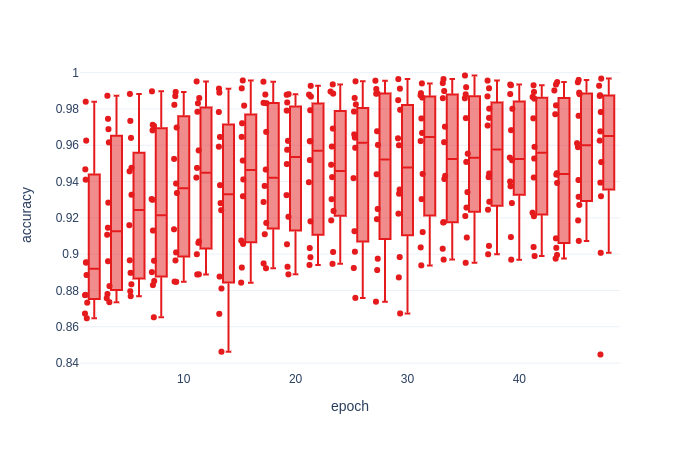}
  \caption{Box-plot showing the variance of the accuracy by epoch - optimiser: sgd. Each point represents a dimension.}
   \label{Fig_Boxplot_Sgd}
\end{figure}

In a nutshell, figs.~\ref{Fig_Majorization_Adam}-\ref{Fig_Majorization_Sgd} show that our models do learn how to classify entanglement transformations with high accuracy. Note, though, that the sgd optimiser performs worse than its contenders - we comment on that apparently undesirable behaviour in the next section. Figs.~\ref{Fig_Boxplot_Adam}-\ref{Fig_Boxplot_Sgd} summarises the variance on the accuracy for each of the optimisers. Again, the sgd optimiser is noticeable worse than the other methods - nonetheless, we emphasise, they all succeed in learning the majorization relation regardless of the dimension.


\begin{figure}
  \includegraphics[scale=0.35]{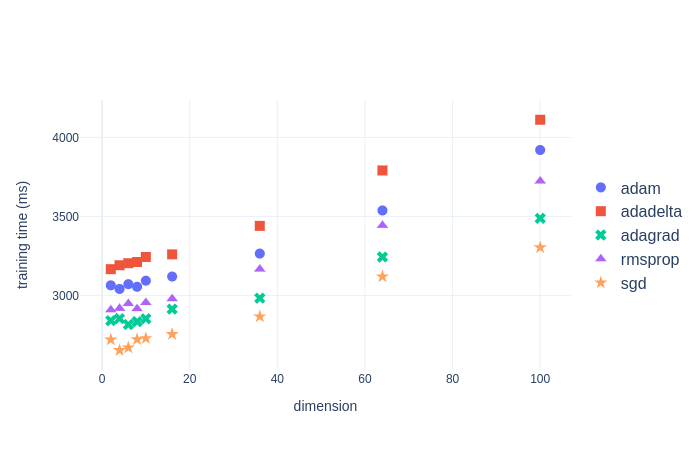}
  \caption{Training time across dimensions by optimiser for 50 epochs}
   \label{Fig_TrainingTime_vs_Dimension}
\end{figure}

\subsection{Machines Learn Self-Catalysis}\label{SubSec.ResultsSelfCatalysis}

The previous section shows that basic machine learning algorithms learn how to detect entanglement transformations with high accuracy. In this sense, one may expect that if we had trained a machine to detect self-catalysis, it would detect such class of transformations within high precision - although no analytical expression is known to dictate this phenomenon.

In our work, we have studied a different situation. The main question we set out to investigate is whether a majorization-trained machine could be used to classify self-catalysis. This section compiles our results for this transfer knowledge task.

\begin{figure}
  \includegraphics[scale=0.35]{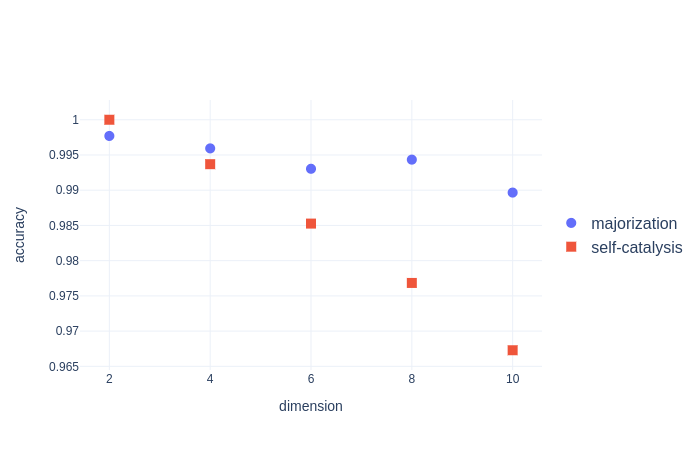}
  \caption{Accuracy variation between majorization and self-catalysis - optimiser: adam.}
   \label{Fig_Delta_Maj_Self_Accuracy_Adam}
\end{figure}

\begin{figure}
  \includegraphics[scale=0.35]{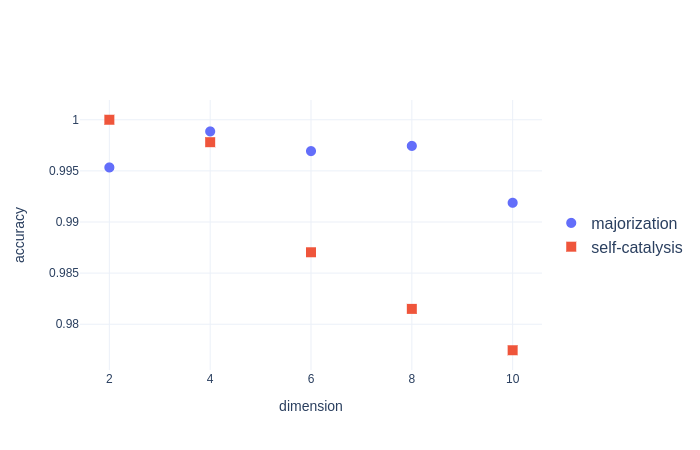}
  \caption{Accuracy variation between majorization and self-catalysis - optimiser: adadelta.}
   \label{Fig_Delta_Maj_Self_Accuracy_Adadelta}
\end{figure}

\begin{figure}
  \includegraphics[scale=0.35]{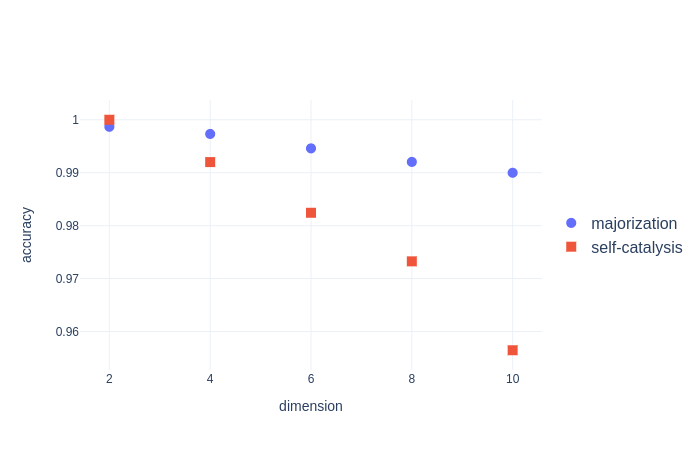}
  \caption{Accuracy variation between majorization and self-catalysis - optimiser: adagrad.}
   \label{Fig_Delta_Maj_Self_Accuracy_Adagrad}
\end{figure}

\begin{figure}
  \includegraphics[scale=0.35]{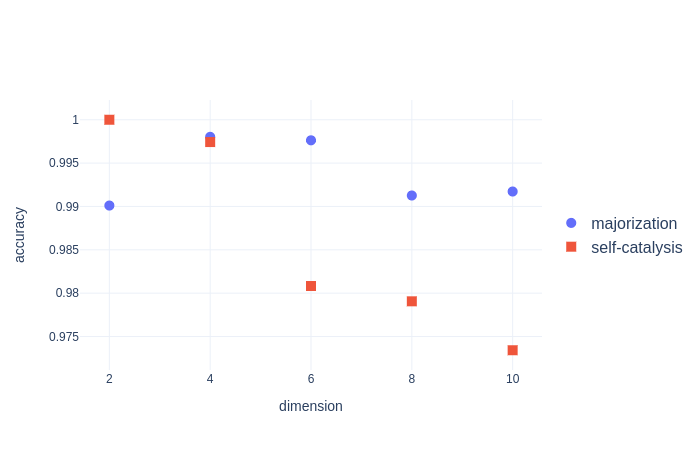}
  \caption{Accuracy variation between majorization and self-catalysis - optimiser: rmsprop.}
   \label{Fig_Delta_Maj_Self_Accuracy_Rmsprop}
\end{figure}

\begin{figure}
  \includegraphics[scale=0.35]{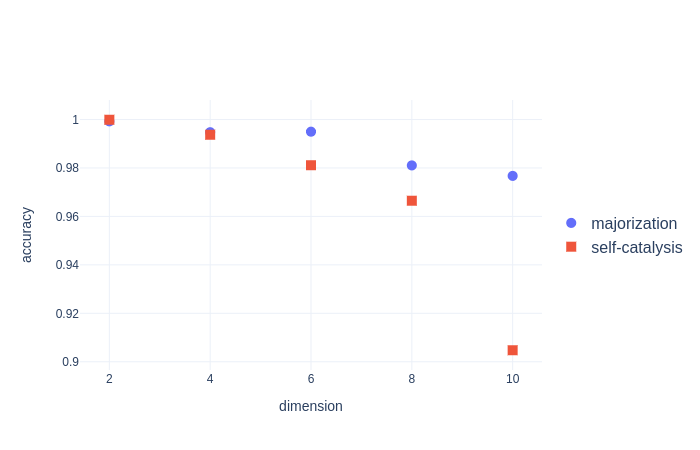}
  \caption{Accuracy variation between majorization and self-catalysis - optimiser: sgd.}
   \label{Fig_Delta_Maj_Self_Accuracy_Sgd}
\end{figure}

Simply put, we have noticed that the accuracy of our machines decreases in this transfer-knowledge task. The difference between the performance is even more latent when we compare the simultaneously the original learning task with the transfer-knowledge situation. Nonetheless, we may safely say that majorization-trained machines still keep a high success rate in identifying self-catalytic transformations (table~\ref{Fig_Table_Self_Catalysis_Accuracy}). Yet again, the sgd optimiser stands out in giving the worst set results. Figs.~\ref{Fig_Delta_Maj_Self_Accuracy_Adam}-\ref{Fig_Delta_Maj_Self_Accuracy_Sgd} display our results for this part. 
\begin{table}
  \includegraphics[scale=0.26]{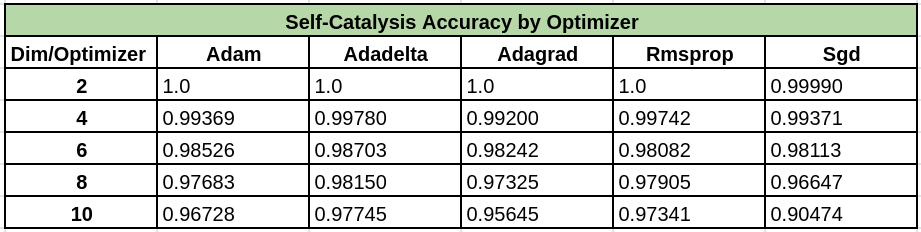}
  \caption{Performance of majorization-trained models for detecting self-catalysis.}
   \label{Fig_Table_Self_Catalysis_Accuracy}
\end{table}
%

\subsection{Machines Learn Higher Order Self-Catalysis}\label{SubSec.ResultsSelfSelfCatalysis}

We have established that classical agents can learn the majorization relation, and that majorization-trained agents can transfer their knowledge two other scenarios - although some agents perform better than others. Now, we want to investigate a hybrid-situation. We want to see if a combination of pre-selection plus our trained machines performs better than a repeated use of trained machines.

More precisely, we have studied two different scenarios. Given two non-interconvertible vectors $\alpha$ and $\beta$, we take $\alpha \otimes \alpha$ and $\alpha \otimes \beta$. That done, we either $(i)$ use the majorization result and compare the accessibility, and for those which are not-interconvertible we feed them into the algorithm in a machine of $d=64$, or $(ii)$ use first the machine for $d=16$ and for those that this machine says are not-interconvertible, we feed them into the machine for $d=64$.

As expected, regardless the strategy, either $(i)$ or $(ii)$ gives excellent accuracy in detecting high-order self-catalysis. If one is willing to admit a small error margin, it is virtually indifferent to squeeze in and check majorization relations in the middle of the process. Put another way, pre-selecting with certainty via thm.~\ref{Thm:NielsenThm} does not bring any noticeable gain. As the calculations involved in using Nielsen's theorem are costly, it is wiser to implement a full machine learning strategy.  Table~\ref{Fig_Table_SelfSelf_Catalysis_Accuracy} summarises this discussion. 
\begin{table}
  \includegraphics[scale=0.2]{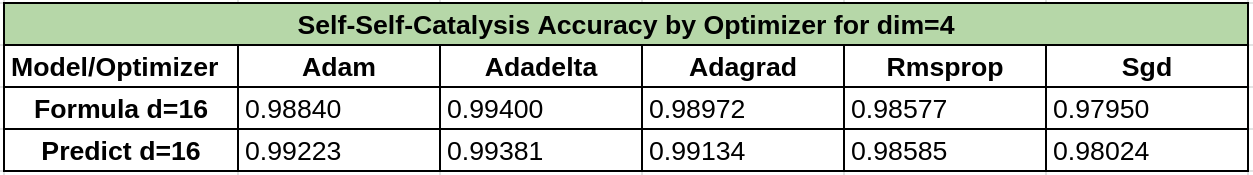}
  \caption{Comparison between the performance of two approaches for detecting higher-order self-catalysis - both using majorization-trained models.}
   \label{Fig_Table_SelfSelf_Catalysis_Accuracy}
\end{table}
%

\section{Conclusions}\label{Sec.Conclusion}

This work has investigated whether a classical agent can learn quantum phenomena through classical data. Naturally, we have modelled the inner functioning of our classical agent using neural networks. The quantum phenomena we have investigated are catalytic and self-catalytic entanglement transformations. We have thoroughly studied test accuracy, test loss, training duration, the network configuration and, finally, different optimisers - so that, to a certain extent, we have evaluated a variety of agents against various figures of merit. Additionally, we have also analysed the performance of an specialised agent in a related but different task. We unpack these results in the following paragraphs.

Firstly and foremost, in this contribution we have exclusively centred our attention at deterministic entanglement transformations involving bipartite states. Those transformations are described by the Nielsen's criterion (Thm.~\ref{Thm:NielsenThm}) and we have seen that classical machine learning algorithms do learn how to detect allowed transformations with fairly high-accuracy (fig.~\ref{Fig_Majorization_Adam} to fig~\ref{Fig_Majorization_Sgd} and fig.~\ref{Fig_Boxplot_Adam} to fig.~\ref{Fig_Boxplot_Sgd}) even in more specialised scenarios of catalysis and self-catalysis (table~\ref{Fig_Table_Self_Catalysis_Accuracy} and table~\ref{Fig_Table_SelfSelf_Catalysis_Accuracy}) - scenarios where there is no general analytical expressions dictating the occurrence of such transformations. Note though that while the other optimisers see an increasing in the accuracy when we increase the training epochs, the stochastic gradient descendent method (sgd) has a large variance in the data (fig.~\ref{Fig_Boxplot_Sgd}). Similarly, sgd also performs poorly in comparison with its fellows optmisers when the dimension gets larger and larger. For self-catalytic transformations, the sgd optimiser does not perform brilliantly either, as table.~\ref{Fig_Table_Self_Catalysis_Accuracy} shows that even for very low dimensions it has doubts of (self-catalytic)transformations that should never happen.  

Secondly, as we mentioned before, we have also compared the performance of our algorithms for slightly different tasks. The question was whether a trained machine for majorization could also identify self-catalysis with high accuracy. In practice, that means that we would not have to start the whole process from scratch for related tasks - reducing costs of time and energy. In theory, that means that our proxy for classical agents could transfer their learning across tasks. Our results are displayed in figs.~\ref{Fig_Delta_Maj_Self_Accuracy_Adadelta}-\ref{Fig_Delta_Maj_Self_Accuracy_Sgd}. Although all optimisers perform fairly well in this cross-task, note that they misidentify some self-catalytic transformations - as they lose accuracy compared to the original task. Curiously, even though the sgd method gets behind all others for higher dimensions, note that it performs very well for low dimensions. This erratic behaviour is an open problem, and we hope to explore it in a follow-up contribution - our ansatz is that we would need a more extensive dataset and many more training epochs for the sgd to reach its sweet spot of accuracy and stability across dimensions.

Thirdly, fig.~\ref{Fig_TrainingTime_vs_Dimension} displays the training time across all dimensions we have considered for each optimiser. The training time grows linearly with the dimension of $\mathbb{V}^{d}$ and that may be used to estimate how long one has to wait for the training of a particular algorithm to be complete. We want to point out that despite giving the most erratic results (within a reasonably good accuracy), the sgd method has the fastest training time, even for higher dimensions. With the cost of sacrificing perfect accuracy, the training time of the algorithms plus the running time of each program turn the investigation of catalytic entanglement transformation realisable in practice - see~\cite{DDT16} for a discussion of the difficulties arising in numerical investigations of that phenomenon. We hope our framework facilitates the numerical study of self-catalytic and catalytic entanglement transformations.    

Although we acknowledge that several other works implement machine learning techniques to tasks in quantum mechanics, particularly classifying entanglement~\cite{HPFP20, CarleoEtAl19, SSP15}, our work differs from the previous ones. Here our objective was three-fold. First, we built our models from scratch. That gave us the ability to change some parameters in the models and monitor the effect of those changes - this is significantly lost when the models are complex or when one uses black-boxes that output some results when input data. Second, we wanted to train a machine and compare the performance of this machine against another task - as if we have trained an agent and ask them to resolve another similar task to verify if learning can be transmitted. Third, we have written this paper as pedagogically as possible and provided all the codes as open source. 

To conclude, in this contribution, we have exclusively focused on deterministic entanglement transformations involving bipartite states. A natural follow-up is the investigation of probabilistic entanglement transformations. In this probabilistic framework, we have a smoother classification, as instead of a 0-1 answer, we allow for any number in the interval $[0,1]$. Yet again, numerical and theoretical investigations are cumbersome, time-consuming or simply virtually impractical in some cases. In this sense, we expect that an analysis similar to what we did here may not only answer "can a classical agent learn a quantum feature?" but it may also reveal new information.  

\begin{acknowledgments}
CD thanks the hospitality of the Institute for Quantum Studies at Chapman University. This research was supported by the National Research, Development and Innovation Office of Hungary (NKFIH) through the Quantum Information National Laboratory of Hungary and through the grant FK 135220. This research was also supported by the Fetzer Franklin Fund of the John E.\ Fetzer Memorial Trust and by grant number FQXi-RFP-IPW-1905 from the Foundational Questions Institute and Fetzer Franklin Fund, a donor advised fund of Silicon Valley Community Foundation.  
 
\end{acknowledgments}

\appendix

\section{Auxiliary Graphs and Tables}\label{Sec.AppGraphs}

Auxiliary graphs and tables can be found on \cite{AD21}.

\section{Gentle Introduction to Neural Networks}\label{Sec.AppIntroduction}
Machine Learning is a sub-field of Artificial Intelligence. It includes algorithms whose main goal is to extract patterns from a \emph{dataset}~\footnote{We are using this terminology to mean the information originated out of the actual collection of data gathered from multiple runs of an experiment as well as an artificially generated set of data. Our work deals mainly with the latter.}, learn from that data and apply the knowledge learned to complete a specific task using similar data, so far never seen by the trained machine \cite{Bishop06,Carbonell90,Bengio09}. There are three types of learning methods namely: unsupervised, supervised and reinforcement learning \cite{Brownlee16}. Each of them has a different set of applications, and it is necessary to understand the nature of the problem one wants to solve to appropriately choose which learning method to implement. In this work, we are interested in classifying pairs of probability vectors. More precisely, we want to determine if they are convertible, catalysis-convertible, or self-catalytic-convertible - convertibility here meaning majorization. For doing so, we use a supervised learning strategy, since classification tasks belong to that learning paradigm~\footnote{In fact, the supervised learning paradigm is subdivided into two types of tasks: classification and regression. The former is tailored to identify and label specific features in a dataset. The latter deals with identify patterns that can be used in predicting future behaviours already congealed in the dataset.}.

One of the forms in which we can implement supervised learning consists in using artificial neural networks. Such networks were initially based on the layered structure of the human neural system. Like a child learning to label things (objects, animals, colors, etc.), the supervised neural network receives inputs with labels attached to them, and after many repetitions and exposure to similar examples, it can learn how to classify a collection of data for what it has been trained for - within a certain accuracy. Following this idea, in this work, we train our neural network with a large set of pair of vectors together with labels, which defines whether it is possible to convert one into the other via (self)catalysis. If this trained model can achieve great results on the tests, we can assume that it has learned how to classify entanglement through majorization.

To organize the concepts and provide an informative summary about neural networks, subsection~\ref{SubSec.NeuralNetworksHistory} gives an overview about its history; subsection~\ref{SubSec.NeuralNetworksArchitecture} explains the neural networks' structure and components definitions, while subsection~\ref{SubSec.NeuralNetworksTraining} details how it works, clarifying the training phase steps.

\subsection{A brief overview of Neural Networks}\label{SubSec.NeuralNetworksHistory}

Comparing boolean logic with the human neural activity involved with decision problems with binary outputs, in 1943, Warren McCulloch and Walter Pitts published the seminal work~\cite{mcculloch43a} that allowed for the development of what now goes by the name of neural networks. It was based on that work that Frank Rosenblatt came up with the Perceptron: the simplest feedforward neural network~\cite{rosenblatt1958perceptron} - portraying a single neuron. Rosenblatt introduced weights into the original protocol, resulting in a linear classifier biologically inspired. To find a better rule on how to update those weights, Paul Werbos was the first one to write about the backpropagation algorithm~\cite{Werbos:74} applied on neural networks. In 1989, Yann LeCun~\cite{LeCunBoserDenkerEtAl89} illustrated how to use a neural network together with a backpropagation logic to train a model that was able to learn how to distinguish handwritten zipcode characters. Since then, many researchers have been exploring and developing new neural networks architectures to solve different problems in the real world.

To give the reader a better understanding of the model we use in our work, we will review a basic feedforward neural network model. The focus is on explaining, as self-contained as possible, its components and how the networks work. We do not intend to give neither a technical review nor an exhaustive account of the ever-growing field of neural networks. For such overviews we suggest~\cite{GBC16} or~\cite{Bishop95}.

\subsection{Neural Networks Architecture}\label{SubSec.NeuralNetworksArchitecture}

The main component of a neural network is a \emph{node}, also called artificial neuron. Nodes are connected to other nodes, and for feedforward networks they can be seen as forming a directed graph with no cycles. Like in a human brain neuron, nodes receive information (signal) from other nodes directly connected with them. Each node combines the received signals, in a weighted sum, and compare the combined value to an activation level. If the threshold is exceeded, an output signal is sent to next nodes connected to its end. Otherwise, this specifically node does not send any data forward.

A feedfoward network is usually composed by one \emph{input layer}, one or more \emph{hidden layers}, and one \emph{output layer}. Nodes on a layer only connect with nodes from the layer ahead, allowing the information to flow only in one direction, forward. A possible configuration consists of having all nodes from the layer \textit{l} connected to all of the other nodes on the $(l+1)$th-layer. This fully connected structure is exactly the configuration we will be using in this work, as shown on fig.~\ref{Fig.NetworkModelConfiguration}.

Each of the nodes's connections comes with an attached \emph{weight} that is adjusted between the iteractions of the training algorithm. Within the supervised learning paradigm, backpropagation provides a way to update those weights based on the network output. Simply put, this update works as follows: in each round and for a given input, the backpropagation algorithm compares the network output to the actual value determined by that given input. It is the difference between the actual value and the network's output value that determines how the weights change across the nodes's connections. We call \emph{learning} the process of weight adjustment to refine the output of the network, based on the desired value, in supervised learning.

With exception of the input layer, where the nodes only get the raw input data and pass it on to the next layer, the nodes from all other layers have a defined \emph{activation function}. It is common to use the same function in all hidden layers, while the output layer typically uses a different one, depending on the type of prediction required by the model~\cite{Reed99}. For networks using backpropagation as a weight updater, we usually require sufficiently smooth activation functions - sigmoid distributions are the most prominent example of activation function, and will be used for the output layer in our work.

There is no receipt to define all the network architecture that will be used for a specific problem. In other words, all network's defining parameters needs to be chosen and tested with a sample dataset, in order to compare the results and select what configuration performs better in each case. This can be an infinite work in terms of possible combinations, so it is usual to select a network already used in the literature for similar problems, and make small changes to it based in your problem.

\subsection{Neural Networks Training}\label{SubSec.NeuralNetworksTraining}

Once the network is defined, the training phase can start. The \emph{dataset} is divided into two sets - the training and test data. This division seeks to avoid that the networks memorizes the entire data instead of learning its pattern~\cite{JWHT13}. Simply put, the training phase is done using a different dataset that is used to evaluate its performance, making the trained model results more generic and prepared for unseen data. Every sample in the training data is codified into the input layer, the subsequent information is passed through the network, layer by layer, until it reaches the output layer, and depending on the task in question this final layer outputs a certain result. The number of times that each sample passes through the network is called \emph{epoch}. Initially, all the nodes' connections' weighs are randomly defined, as they will be adjusted right after the output layer's result. It is here that the backpropagation comes into place. Starting from the output layer, the algorithm goes backwardly adjusting the weights of each connection. Roughly speaking, this adjustment is based on the difference between the obtained result and the desired value for a given input. After the training halts - because the maximum number of epochs has been achieved, or for another stopping condition~\cite{Bishop06,GBC16} - the final configuration of weights and parameters determines the trained model. The trained model can be used to predict the output of samples in the test dataset. This output, when compared to the actual value for this sample, gives the \emph{performance} of this trained model.

For an in-depth and pedagogical explanation of our exposition above, we suggest the example in ref.~\cite{Veloso21}, where the author works with a \emph{Multi-layer Perceptron} (MLP) containing a single hidden layer and uses backpropagation to update its weights.

\section{Codes}\label{Sec.AppCodes}

The pseudocode we have used to generate our dataset is displayed in the Algorithm~\ref{alg:generate_samples} below. To facilitate the reading, the other codes we have used in this work can be found on \cite{AD21}.

All experiments were made using an Asus laptop with Intel® Core™ i7-8550U CPU @ 1.80GHz × 8, 16GB RAM, 512GB SSD, NV138 / Mesa Intel® UHD Graphics 620 (KBL GT2) and Fedora 32 (Workstation Edition) 64-bit.

\begin{widetext}
\begin{center}
\begin{algorithm}[H]
\caption{Generate Dataset}\label{alg:generate_samples}
\begin{algorithmic}[1]
\Require $dim \geq 0$, $n \geq 0$
\Ensure $csv\_file \gets alpha\_n[i];beta\_n[j];maj\_alpha\_beta[i][j];maj\_beta\_alpha[j][i]$, $\forall\,\, 0 \leq i \leq n$, $0 \leq j \leq n$
\State $alpha\_n \gets generate\_random\_sorted\_vector(dim,n)$
\State $beta\_n \gets generate\_random\_sorted\_vector(dim,n)$

\Comment{$generate\_random\_sorted\_vector(dim,n)$ generates $n$ vectors of size $dim$ with uniform probability distribution, sorted descendent}
\State $i \gets 1$
\State $j \gets 1$
\State $maj\_alpha\_beta \gets [\,\,]$
\State $maj\_beta\_alpha \gets [\,\,]$
\While {$i \leq n$}
\While {$j \leq n$}
\State $maj\_alpha\_beta[i][j] \gets check\_rule(alpha\_n[i],beta\_n[j])$
\State $maj\_beta\_alpha[i][j] \gets check\_rule(beta\_n[i],alpha\_n[j])$
\EndWhile
\EndWhile
\end{algorithmic}
\end{algorithm}
\end{center}
\end{widetext}

\bibliography{biblio}
\end{document}